# The quantum mechanics correspondence principle for spin systems and its application for some magnetic resonance problems

Victor Henner[1,3,a], Andrey Klots[2,b], Tatyana Belozerova[3]

[1]Department of Physics and Astronomy, University of Louisville, KY, 40292, USA

[2]Department of Physics and Astronomy, Vanderbilt University, TN, 37240, USA

[3]Department of Physics, Perm State University, Perm, 614990, Russia

[a]vkhenn01@louisville.edu, [b]klotsandrey@gmail.com

**Abstract**

Problems of interacting quantum magnetic moments become exponentially complex with increasing number of particles. As a result, classical equations are often used but the validity of reduction of a quantum problem to a classical problem should be justified. In this paper we formulate the correspondence principle, which shows that the classical equations of motion for a system of dipole interacting spins have identical form with the quantum equations. The classical simulations based on the correspondence principle for spin systems provide a practical tool to study different macroscopic spin physics phenomena. Three classical magnetic resonance problems in solids are considered as examples – free induction decay (FID), spin echo and the Pake doublet.

## 1. Correspondence principle for interacting magnetic moments

The correspondence between the Heisenberg equation $\dot{\hat{\vec{\mu}}} = -i/\hbar \left[ \hat{\vec{\mu}}, \hat{\mathcal{H}} \right]$ for operator $\hat{\vec{\mu}}$ of magnetic moment in an external field $\vec{H}$ (the Hamiltonian is $\hat{\mathcal{H}} = -\hat{\vec{\mu}}\vec{H}$) and classical equation of motion $\frac{d\vec{\mu}}{dt} = \gamma \vec{\mu} \times \vec{H}$ ($\gamma$ is the gyromagnetic ratio) is obvious when individual magnetic moments $\vec{\mu}_k$ do not interact with each other. But does this correspondence take place in case of interacting magnetic moments and in the presence of other factors? In fact, in this section we will formulate the analog of Ehrenfest's theorem for a system of dipole-interacting magnetic moments. The validity of such correspondence justifies modeling the dynamics of the expectation values of quantum observables using the classical equations.

In the *quantum description* consider a spin system described by the Hamiltonian

$$\hat{\mathcal{H}} = \hat{\mathcal{H}}_Z + \hat{\mathcal{H}}_{dd}. \tag{1}$$

The Zeeman part of the Hamiltonian

$$\hat{\mathcal{H}}_Z = -\sum_k \hat{\vec{\mu}}_k \vec{H} = -\hbar \sum_k \gamma_k H^\alpha \hat{S}_k^\alpha \tag{2}$$

gives the interaction with an external classical magnetic field $\vec{H}$, $\hat{\vec{\mu}}_k = \gamma_k \hbar \hat{\vec{S}}_k$ are operators of magnetic moments, $\hat{S}_k^\gamma$ are dimensionless spin operators, $\gamma_k$ is the gyromagnetic ratio of $k$-th particle. To consider particles or clusters with different spins we allow the possibility of different gyromagnetic ratios by assigning the index $k$ in $\gamma_k$. Latin letters represent particle numbers and Greek letters – spatial dimensions. Also, summation over repeating Greek indexes is implied and the external field can depend on position and time.

The dipole Hamiltonian is

$$\hat{\mathcal{H}}_{dd} = \frac{\hbar^2}{2} \gamma_l \gamma_k \sum_{l \neq k} D_{lk}^{\alpha\beta} \hat{S}_l^\alpha \hat{S}_k^\beta, \quad (3)$$

where

$$D_{lk}^{\alpha\beta} = \frac{1}{r_{lk}^3} \delta_{\alpha\beta} - \frac{3}{r_{lk}^5} r_{lk}^\alpha r_{lk}^\beta. \quad (4)$$

In the quantum description based on the Heisenberg equation (over dots in this and the following equations indicate the time derivative)

$$\dot{\hat{S}}_l^\delta = \frac{-i}{\hbar} \left[ \hat{S}_l^\delta, \hat{\mathcal{H}}_Z + \hat{\mathcal{H}}_{dd} \right].$$

The contribution of $\hat{\mathcal{H}}_Z$ in $\dot{\hat{S}}_l^\delta$ is $i \sum_j \gamma_j H^\alpha \left[ \hat{S}_l^\delta, \hat{S}_j^\alpha \right]$ and with $\left[ \hat{S}_l^\delta, \hat{S}_j^\alpha \right] = i \delta_{lj} e_{\beta\delta\alpha} \hat{S}_l^\beta$ this term results in

$$\dot{\hat{S}}_l^\delta = \gamma_l e_{\delta\beta\alpha} \hat{S}_l^\beta H^\alpha. \quad (5)$$

For operators $\hat{\vec{\mu}}_l = \gamma_l \hbar \hat{\vec{S}}_l$ equation (5) becomes

$$\dot{\hat{\vec{\mu}}}_l = \gamma_l \hat{\vec{\mu}}_l \times \vec{H}. \quad (6)$$

The contribution of dipolar interactions in $\dot{\hat{S}}_l$ is $-i \left[ \hat{S}_l^\delta, \hat{\mathcal{H}}_{dd} \right] / \hbar$ and with $[a, bc] = [a, b]c + b[a, c]$ we obtain

$$\dot{\hat{S}}_l^\delta = \hbar \frac{1}{2} \left( \sum_k \gamma_l \gamma_k e_{\delta\alpha\sigma} D_{lk}^{\alpha\beta} \hat{S}_l^\sigma \hat{S}_k^\beta + \sum_j \gamma_j \gamma_l e_{\delta\beta\sigma} D_{jl}^{\alpha\beta} \hat{S}_j^\alpha \hat{S}_l^\sigma \right).$$

Since tensor $D_{lk}^{\alpha\beta}$ is symmetric in lower and upper indexes, in the second term we can interchange $\alpha$ and $\beta$, and change the summation index $j$ to $k$ which gives

$$\dot{\hat{S}}_l^\delta = \hbar \gamma_l \frac{1}{2} \sum_k \gamma_k e_{\delta\alpha\sigma} D_{lk}^{\alpha\beta} (\hat{S}_l^\sigma \hat{S}_k^\beta + \hat{S}_k^\beta \hat{S}_l^\sigma)$$

and since $l \neq k$

$$\dot{\hat{S}}_l^\delta = \gamma_l \sum_k e_{\delta\alpha\sigma} D_{lk}^{\alpha\beta} \left( \gamma_k \hbar \hat{S}_k^\beta \right) \hat{S}_l^\sigma. \quad (7)$$

For operators $\hat{\vec{\mu}}_l = \gamma_l \hbar \hat{\vec{S}}_l$ equation (7) becomes

$$\dot{\hat{\mu}}_l^\delta = \gamma_l \sum_k e_{\delta\alpha\sigma} D_{lk}^{\alpha\beta} \hat{\mu}_k^\beta \hat{\mu}_l^\sigma. \quad (8)$$

Combining (6) and (8) we arrive at the equation of motion for the operator of the magnetic moment in both external and dipole magnetic fields:

$$\dot{\hat{\mu}}_l^\delta = \gamma_l \sum_k e_{\delta\alpha\sigma} \left( \hat{\mu}_l^\alpha H^\sigma + D_{lk}^{\alpha\beta} \hat{\mu}_k^\beta \hat{\mu}_l^\sigma \right). \quad (9)$$

In a *classical description* the dynamics of vector $\vec{\mu}_l$ is given by equation

$$\dot{\vec{\mu}}_l = \gamma_l \vec{\mu}_l \times (\vec{H} + \vec{H}_l), \quad (10)$$

where $\vec{H}$ is an external field, $\vec{H}_l$ is the magnetic (dipole) field at the coordinate of the $l$-th magnetic moment.

The dipole energy of a system of classical magnetic moments is

$$E_{dd} = \frac{1}{2} \sum_{l \neq k} D_{lk}^{\alpha\beta} \mu_l^\alpha \mu_k^\beta. \qquad (11)$$

This gives the field $\vec{H}_l$

$$H_l^\alpha = -\partial E_{dd} / \partial \mu_l^\alpha = -\frac{1}{2} \sum_k \left( D_{lk}^{\alpha\beta} \mu_k^\beta + D_{kl}^{\alpha\beta} \mu_k^\beta \right) = -\sum_k D_{lk}^{\alpha\beta} \mu_k^\beta. \qquad (12)$$

Therefore, equation (10) becomes

$$\dot{\mu}_l^\delta = \gamma_l e_{\alpha\sigma\delta} \left( H^\alpha + H_l^\alpha \right) \mu_l^\sigma = \gamma_l \sum_k e_{\delta\alpha\sigma} \left( \mu_l^\alpha H^\sigma + D_{lk}^{\alpha\beta} \mu_k^\beta \mu_l^\sigma \right). \qquad (13)$$

The effect of the exchange interactions, potentially important for electron spin systems, can be taken into account by adding to the Hamiltonian the exchange term

$$\hat{\mathcal{H}}_J = \frac{1}{2} \sum_{l \neq k} J_{lk} \hat{\vec{S}}_l \hat{\vec{S}}_k. \qquad (14)$$

It has the same spin structure as $\hat{\mathcal{H}}_{dd}$, therefore the exchange term can be included in both the equations (9) and (13) as the dipole one just with the coefficients $J_{lk}$ instead of tensors $D_{lk}^{\alpha\beta}$.

Equations (9) and (13) give the correspondence between the classical and quantum pictures for interacting magnetic moments: the Heisenberg equation for operators $\hat{\vec{\mu}}_l$ and classical equation (10) for vectors $\vec{\mu}_l$ result to equations identical in form. This means that, the dynamics of the expectation values of quantum operators, $\langle \hat{\vec{\mu}}_l \rangle$, should be similar to dynamics of the corresponding classical quantities, $\vec{\mu}_l$.

Note, that even though the correspondence we have proved seems to be intuitively expected, it is not obvious in advance. Moreover, despite being valid for spin components $\hat{\vec{S}}_k$ and $\vec{S}_k$ (as well as for components of the total spin $\sum_k \hat{\vec{S}}_k$ and $\sum_k \vec{S}_k$) it is not valid for some other observables, such as the magnitude of the total spin – it can be shown that for dipole interacting spins the equations for $\frac{d}{dt} \left( \sum_k \hat{\vec{S}}_k \right)^2$ and $\frac{d}{dt} \left( \sum_k \vec{S}_k \right)^2$ are different. Hence, the correspondence principle we obtained is more specific than a general principle that quantum equations of motion in the classical limit $\hbar \to 0$, $S \to \infty$ coincide with the macroscopic ones.

The principle discussed above gives the physical base for often used modeling of spin systems dynamics with classical equations. It is one of the unique cases when a general Erenhfest's theorem, which states that the classical mechanics Hamilton equations hold for operators expectation values result in the practical tool. Surprisingly, this correspondence principle for dipole interacting spins was never discussed and formulated earlier, to the best of our knowledge.

Clearly, the classical approach cannot provide the energies of individual states and the transitions between them. But when the macroscopic quantities characterizing the spin system as a whole are evaluated, the examples below show that the equations for individual spins, formally coinciding with the quantum ones, nicely describe some fundamental NMR features. We find that in modeling this collective phenomenon the number of spins necessary to provide reliable results should be at least several hundred. No quantum 3D calculations are feasible for systems with $2^N$ states for this large a value of $N$.

## 2. Three applications of correspondence principle: Free induction decay, Spin echo and the Pake doublet

Fist, let us formulate the equations of motion in the form convenient for simulation of spin dynamics problems. As always, it is useful to split the dipole Hamiltonian into secular $\hat{\mathcal{H}}_{dd}^s$ and non-secular $\hat{\mathcal{H}}_{dd}^{ns}$ components:

$$\hat{\mathcal{H}}_{dd}^s = \frac{\hbar^2}{2}\sum_{l\neq k}\gamma_l\gamma_k a_{lk}\left[\hat{S}_l^z\hat{S}_k^z - \frac{1}{4}(\hat{S}_l^+\hat{S}_k^- + \hat{S}_l^-\hat{S}_k^+)\right], \qquad (15)$$

$$\hat{\mathcal{H}}_{dd}^{ns} = \frac{\hbar^2}{2}\sum_{l\neq k}\gamma_l\gamma_k\left[b_{lk}\hat{S}_l^+\hat{S}_k^+ + b_{lk}^*\hat{S}_l^-\hat{S}_k^- + 2c_{lk}\hat{S}_l^+\hat{S}_k^z + 2c_{lk}^*\hat{S}_l^-\hat{S}_k^z\right], \qquad (16)$$

where the coefficients are

$$a_{lk} = D_{lk}^{zz}, \quad b_{lk} = \frac{1}{4}(D_{lk}^{xx} - D_{lk}^{yy} - 2iD_{lk}^{xy}), \quad c_{lk} = \frac{1}{2}(D_{lk}^{xz} - iD_{lk}^{yz}). \qquad (17)$$

In contrast to $\hat{\mathcal{H}}_{dd}^s$, $\hat{\mathcal{H}}_{dd}^{ns}$ does not commute with the z-component of the total magnetic moment.

Let us define the angular frequencies related to the external constant field $\vec{H}_0$ (its direction is taken for the *z* axis), mean value of dipolar field, $H_d = \mu/a^3$, and their ratio:

$$\omega_0 = |\gamma|H_0, \quad \omega_d = |\gamma|H_d, \quad p_d = \omega_d/\omega_0, \qquad (18)$$

where *a* is the mean distance between adjacent voxels (in simulations it is the lattice parameter of the cubic unit cell) and $\mu$ is a mean value of the modulus of a magnetic moment (in simulations below we consider particles with same spin). Notice, that $\omega_d$ is just a characteristic of the mean value of the local dipole field ($\omega_d$ is $\omega_{loc}$), the actual dipole fields at the locations of each spin are calculated by equation (12).

Let us write the classical equation (13) for the magnetic moment $\vec{\mu}_l$ of the *l*-th particle in the form

$$\frac{d}{dt}\vec{\mu}_l = |\gamma|\vec{\mu}_l \times \left(\vec{H}_0 + \vec{H} + \vec{H}_l\right). \qquad (19)$$

Here $\vec{H}_0 = (0, 0, H_0)$ is an external longitudinal field, $\vec{H} = (H^x, H^y, 0)$ is transverse field, and $\vec{H}_l = (H_l^x, H_l^y, H_l^z)$ is the dipole field at the location of the *l*-th spin, determined by equation (12). It is convenient to use dimensionless dipole field $\tilde{\vec{H}}_l$ and unit vectors of magnetic moments $\vec{e}_l$ for the *l*-th spin, dimensionless transverse fields $h^{x,y}$, and a dimensionless time $\tilde{t}$ given by

$$\vec{H}_l/H_0 = p_d\tilde{\vec{H}}_l, \quad \vec{e}_l = \vec{\mu}_l/\mu, \quad h^{x,y} = H^{x,y}/H_0, \quad \tilde{t} = \omega_d t. \qquad (20)$$

Using these definitions, we obtain equation (19) in the form:

$$\begin{cases} \dfrac{de_l^x}{d\tilde{t}} = (e_l^y - e_l^z h^y)/p_d + \left(e_l^y\tilde{H}_l^z - e_l^z\tilde{H}_l^y\right), \\[4pt] \dfrac{de_l^y}{d\tilde{t}} = (-e_l^x + e_l^z h^x)/p_d + \left(-e_l^x\tilde{H}_l^z + e_l^z\tilde{H}_l^x\right), \\[4pt] \dfrac{de_l^z}{d\tilde{t}} = \left(e_l^x h^y - e_l^y h^x\right)/p_d + \left(e_l^x\tilde{H}_l^y - e_l^y\tilde{H}_l^x\right). \end{cases} \qquad (21)$$

Let us start with the free-induction decay (FID) of the transverse magnetization in a dipolar-coupled rigid lattice – a fundamental problem in magnetic resonance and in the theory of many-body interactions [1]. In this situation the Hamiltonian consists of the Zeeman $\hat{\mathcal{H}}_Z = -\sum_k \hat{\mu}_k^z H_0$ and dipole $\hat{\mathcal{H}}_{dd}$ parts (no terms with $h^{x,y}$ in equations (21)). First time similar simulation of FID for classical spins was performed by Jensen and Platz [2] for each spin interacting with 26 neighbors.

The time evolution of the NMR signal is obtained by numerically integrating the system of $3N$ equations (21). Since the modulus of individual moment is conserved, we exploited this restraint to control the stability of the integrations. For the same purpose three different algorithms, Runge-Kutta, Runge-Kutta-Feldberg, and Dormand and Prince were used to make sure they give same results.

Figure 1 presents the results of our simulations for a cubic sample with a thousand spins ($N_x = N_y = N_z = 10$, the periodic boundary conditions are imposed) all coupled by dipolar interactions. Panel (a) demonstrates the transverse polarization, $e^x(\tilde{t}) = (1/N)\sum_l e_l^x(\tilde{t})$, its initial value for simulation in Fig.1 is taken $e^x(0) = 0.70$. After an initial steady part amplitude suddenly decreases and vanishes. The decays are non-exponential and can be characterized by the signal half-life, which in Fig.1a is about $\tilde{t} \approx 1.5$ (in units $a^3/|\gamma|\mu$) – similar to results obtained in [2]. A very remarkable feature of the decay curves is their oscillatory behavior with the characteristic time of about $1/\omega_d$.

It may be more illustrative to present the time scale in physical units for particular nuclei. Consider, for example, a spin system of protons with the mean value of the local dipole field of several gauss, let say, $H_d \approx 2.5G$. In this case the value $\tilde{t} = 1$ corresponds to $t \approx 14\mu s$ (this time doubles for $H_d \approx 1.25G$, etc.) As can be seen in Fig.1a, with such a value $H_d$, the signal half-time is about $T_{1/2} \approx 20\mu s$ and the signal vanishes at about $\tilde{t} \approx 2$, which corresponds to $t \approx 30\mu s$.

Function $e^x(t)$ is proportional to the amplitude of the free precession signal, $G(t)$, its Fourier transform is the shape function $f(\omega)$ [1], meaning that experimental observations of $G(t)$ and $f(\omega)$ are complementary. The Fourier transform (FT) of $e^x(\tilde{t})$ is shown in Fig.1b. The spectrum centered at $\omega_0$ (in dimensionless units $\tilde{\omega} = \omega/\omega_0$ it is $\tilde{\omega}_0 = 1$) has a width (the broadening) determined by the mean value of the dipole field, $H_d$, which in dimensionless units is given by parameter $p_d$ (same dependence on $p_d$ is seen in Fig.4b,d). For spin system of protons with $H_d = 2.5G$ in external field $H_0 = 1T$ (the parameter $p_d = 0.25 \cdot 10^{-3}$), the frequency broadening in Fig.1b is about $\Delta f \sim \gamma H_d / 2\pi \sim 10 kHz$.

Panel (c) shows that Abragam's trial function [1] $f(t) = 0.7 e^{-\frac{a^2 t^2}{2}} \frac{\sin bt}{bt}$ fits $e^x(\tilde{t})$ pretty well. The ratio $b/a$ is close to the ratio of the moments $M_4/M_2^2$ evaluated by Van Vleck [3].

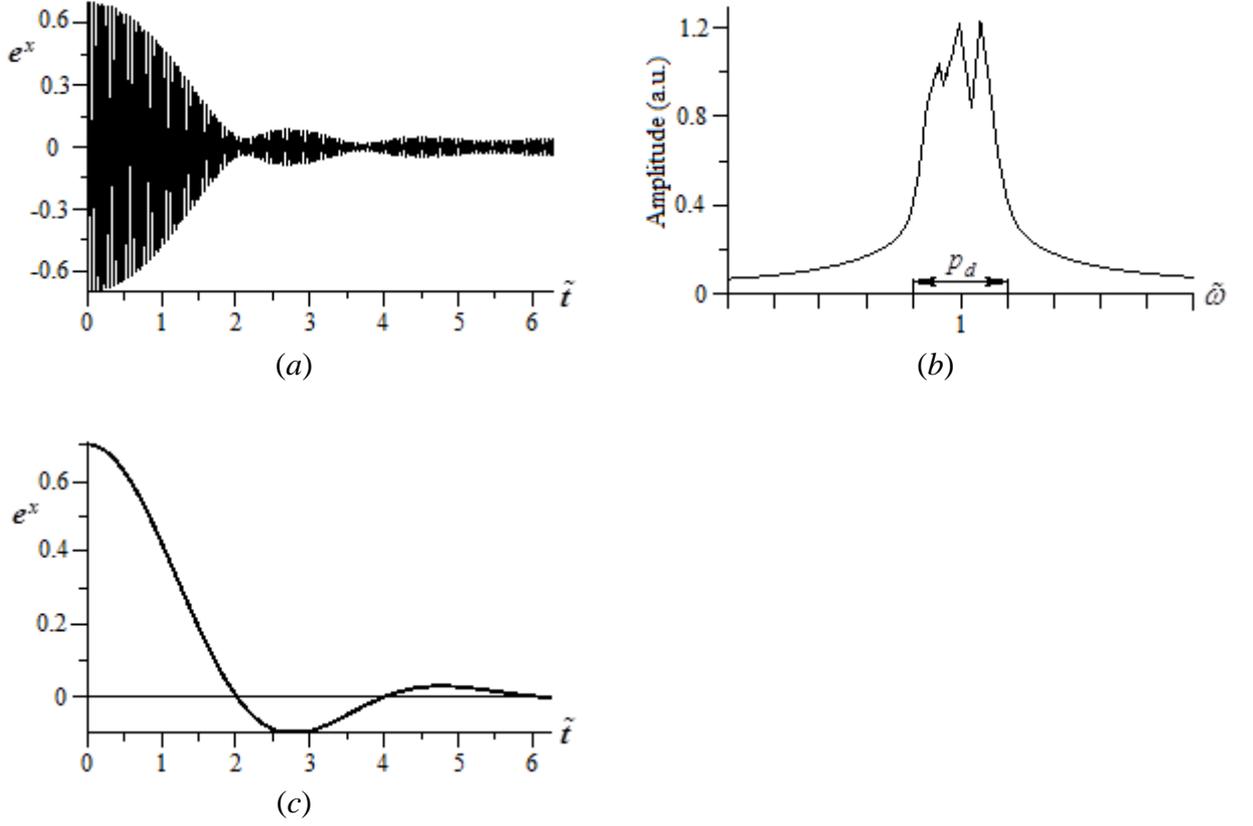

Fig.1. Free precession signal $G(t)$ and dipolar shape function $f(\omega)$;
(a) graph of $e^x(\tilde{t})$ for $e^x(0) = -0.7$;
(b) shape function $f(\omega)$, the amplitude is in arbitrary units;
(c) Abragam's trial function $f(t)$ with parameters $a = 0.00244$, $b = 0.0122$.

Now consider spin echo. First, let us show that the correspondence between the quantum and classical equations remains valid in an arbitrary rotating frame. Transformation to a rotating frame for classical vectors $S_l^\beta$ is performed with the 3D rotation tensor, $S_l'^\alpha = R^{\alpha\beta} S_l^\beta$, followed by substitution $S_l'^\alpha$ instead of $S_l^\alpha$ in the equations of motion (13) for magnetic moments. In the quantum case, transformation of the operators $\hat{S}_l^\alpha$ to a rotating frame is performed using a unitary rotation operator $\hat{U}$, $\hat{S}_l'^\alpha = \hat{U}^+ \hat{S}_l^\alpha \hat{U}$. For, example, for rotation around the $\vec{n}$ axis, $\hat{U} = \exp(i\hat{\vec{S}}\vec{n}\omega t)$, where $t$ is the pulse time with the angular frequency $\omega$. The transformation with operator $\hat{U}$ can be also expressed as $\hat{S}_l'^\alpha = \hat{U}^+ \hat{S}_l^\alpha \hat{U} = R^{\alpha\beta} \hat{S}_l^\beta$ with the same tensor $R^{\alpha\beta}$ as in classical case. Replacement of operators $\hat{S}_l^\alpha$ by $\hat{S}_l'^\alpha$ in the equation of motion (7) (since $l \neq k$ this equation does not contain non-commuting operators) transforms it to the form exactly the same as the equations for classical spin vectors $S_l'^\alpha$. This means that correspondence between the classical and quantum equations of spin motion holds in a rotating frame as well. This result is valid for the complete dipole Hamiltonian (3) (as well as for the exchange Hamiltonian (14)), but only for its secular part when switching to the frame rotating with the frequency $\omega_0$ allows to get rid of fast Larmor oscillations.

If operator $\hat{U}$ consists of $N$ consequent rotations, $\hat{U} = \hat{U}_N...\hat{U}_2\hat{U}_1$, then tensor $R$ is a product of corresponding tensors, $R = R_N...R_2R_1$. Consequent rotations are widely used to obtain different types of spin echo. By applying a suitable sequence of strong rf fields, a system of dipolar-coupled nuclear spins can be made to behave as though the sign of the secular dipolar Hamiltonian had been effectively reversed and the system then appears to develop backward in time. In solids, a perfect refocusing of the free induction decay was obtained (famously) in the "magic sandwich" echo experiment of Rhim et al. [4]. The sequence of unitary rotations used in [4] effectively transforms the secular dipolar Hamiltonian $\hat{\mathcal{H}}_{dd}^s$ to $-k\hat{\mathcal{H}}_{dd}^s$ (with $k$ lies between 1 and 1/2) at time $\tau$ of about several free induction decay time. Because of a formal equivalence of quantum and classical equations in respect to any sequence of rotations, the corresponding reverse $t \to -kt$ at time $\tau$ can be made in classical equations (21), equivalent multiplication of the right sides of these equations by $-k$. This result is another useful consequence of the correspondence principle presented in this study. For spin dynamics simulations this allows us to avoid numerically very complicated multiple - non trivial changes of the initial conditions after specific rotations (corresponding to rf impulses) for evolving spins. If the classical approach adequately describes FID (as we demonstrated above), this time reversal makes the results of spin echo simulations expectable, but it provides a further possibility to check the applicability of classical equations for simulations of spin dynamics in situations when straightforward quantum simulations involving individual spins are not feasible.

Figure 2 shows the results of spin echo simulations for a cubic sample of $N = 10 \times 10 \times 10 = 1000$ spins with almost all (98%) initially directed along the $x$ axis, similar to that prepared after $\pi/2$ pulse (in the simulations in Fig.1 an initial polarization of 70% was used to provide a variety of results). The $x$ component of the magnetization in a constant field $H_0$ after the initial steady part decays to zero because of dephasing of spins coupled by dipole interactions (here the secular dipole Hamiltonian is considered and simulations are performed in a frame rotating with the Larmor), followed by small restorations, similar to Fig.1. Then, at time $\tau = \tilde{t} = 5$ (for protons and the characteristic dipole field $H_d = 2.5G$ this time is about $t \approx 70\mu s$) - the change $t \to -t$ (for simulations in panel (a)), and $t \to -t/2$ (in panel (b)) is made in the equations during the simulation process. A perfect echo appears in both cases – in case (b) it begins at a time twice as large as that in case (a); its duration is also twice that shown in panel (a). These results nicely illustrate the magic spin echo phenomena in solids.

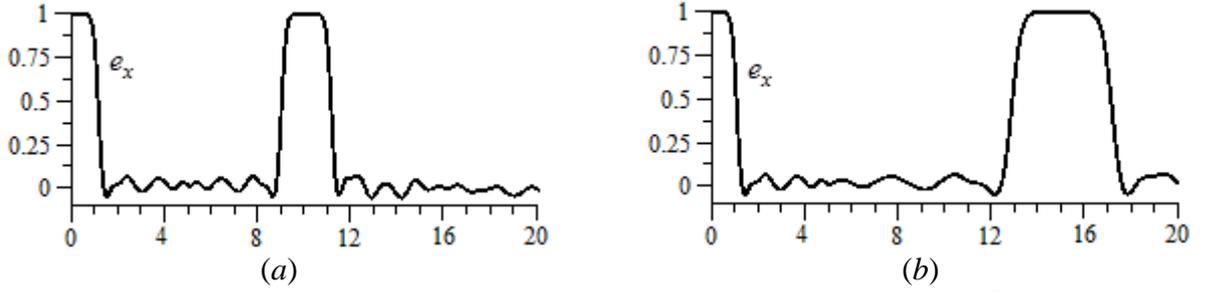

Fig.2. Spin echo simulations. Panel (a) – time reversal $t \to -t$ is made at time $\tilde{t} = 5$; panel (b) – time reversal $t \to -t/2$ is made at time $\tilde{t} = 5$.

Now, after demonstrating the capability of the approach based on the correspondence with the quantum equations to describe the macroscopic phenomena of spin physics, the remaining question is that how many spins are needed for reliable simulations. In order to make the estimations more transparent we consider 1D simulations and arrange spins in line along the z-axis (in this case the non-secular dipole Hamiltonian vanishes) and consider different number of spins. In Figure 3 FID curves for $e^x(\tilde{t}) = (1/N) \sum_l e_l^x(\tilde{t})$ are presented for 2, 25, 50 and 1500 spins (all the figures are filled with the Larmor oscillations with frequency $\omega_0$). For two spins the periodic behavior with the expected period $2\pi$ (in units of $\tilde{t}$) is observed. For several tens of spins, evidence for the collective decay already appears. For number of spins more than 100 the decay time practically does not change with increasing of N, but the oscillations (the "tail") become unchanged only for $N \sim 1000$. Thus, we can conclude that a number of spins for a reliable simulations of macroscopic phenomena is about 1000 or more. We use parallel computations on graphics clusters with 12x240 nodes allowing simulations with several thousand spins.

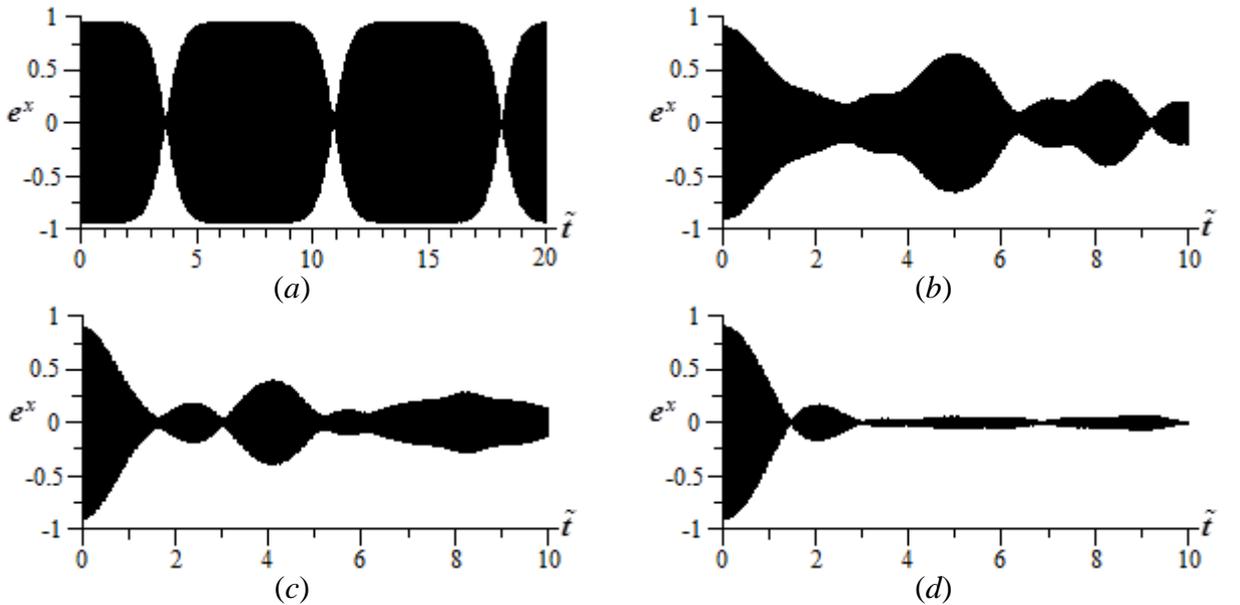

Fig.3. Graph of $e^x(\tilde{t})$ for spins in line along the z-axis. (a) 2 spins, (b) 25 spins, (c) 50 spins, (d) 1500 spins.

Notice here a possibility of quantum simulations for large spin systems with algorithm SPINACH [5], by exclusion the states this algorithm considers to be unimportant and unpopulated. The FID in this approach can be described for 1D case, in some instances 2D simulations can be made, but for 3D situation the understanding is sketchy and simulations prohibitively hard, even in reduced state spaces [6].

Next, consider the Pake doublet - a characteristic absorption line shape in solid-state NMR observed for different directions of external field $H_0$ [7]. Qualitatively it can be explained in the following way. Each nuclear spin produces at its partner a field of several gauss, the component of which along the external constant field $H_0$ alter somewhat the effective large field. If $\vartheta$ is the angle between $H_0$, taken in the $z$ direction, and the line joining the two interacting nuclei, the magnitude of the effective field at one nucleus of the pair can be (for spin ½), in a simple way, written [7] as $H_{eff} = H_0 \pm \alpha(3\cos^2\vartheta - 1)$, where the $\pm$ sign attempts to account for the two possible values of the $z$ component of the partner's magnetic moment, and $\alpha$ is an interaction field parameter (characteristic the local dipole field $H_d$). The field $H_{eff}$ can be presented in equivalent form as $H_{eff} = H_0\left[1 \pm p_d(3\cos^2\vartheta - 1)\right]$. This naive picture predicts a pair of nuclear resonance lines symmetrically disposed about the Larmor frequency in field $H_0$.

To demonstrate that the results of the classical approach describe well the Pake phenomena, we consider 2 and 100 spins forming a line orientated at different angles relative to the field $H_0$. When the line is at the "magic" angle $\vartheta_{magic} = \arccos(1/\sqrt{3}) \approx 54.7^o$ with the Oz axis, the secular interactions vanish, while when the line is along the $Oz$ axis, the non-secular interactions vanish. However, when the line is along the $Ox$ axis, neither the secular interactions nor the non-secular interactions are zero. Panels from (a) to (d) of Fig.4 give the frequencies for a system of two spins, while panels (e) and (f) provide frequencies for 100 spins. For spins placed in a line (1D case), the features of the Pake doublet can be more obviously demonstrated. For a three dimensional case the Pake phenomena is seen as the fine structure in Fig.1b with the frequency spread $\omega_d \sim \gamma\mu/a^3$ (which corresponds to $p_d$ in dimensionless units).

The results in Fig.4 are in a remarkable agreement with the Pake phenomena. When the main (in a strong field) secular dipole Hamiltonian is absent, there is no frequency split (panel (a)). The splitting in panel (b) is close to the value $p_d$, in panel (c) it is about half of this value. In order to show that the splitting is determined by the value of $p_d$, in panel (d) the value of $p_d = 0.001$ is 10 times less than that in panel (b), correspondingly the frequency splitting is 10 times less. Notice, that for three spins the number of peaks for cases (b) and (c) doubles. Also notice, that large values of the parameter $p_d$ were used here just for presentation purposes – the horizontal scale in Fig.4 is proportional to $p_d$. For 100 spins the results are very similar: in panel (e) for the angle $\vartheta_{magic}$ there is no energy split, in panel (f) the energy split is about $p_d = 0.05$. For a line of 100 spins along the $Ox$ axis (not shown), the frequency spread is about two times less than in panel (f). The non-secular dipole Hamiltonian leads to weak additional peaks at zero and double Larmor frequency (to see it better, in panel (e) bigger value of $p_d$ is taken). This is in agreement with the fact that in quantum perturbation theory $\hat{\mathcal{H}}'_{dd}$ determines the energy levels of each spin in the effective fields of other spins, whereas $\hat{\mathcal{H}}''_{dd}$ gives transitions between those energy levels [1].

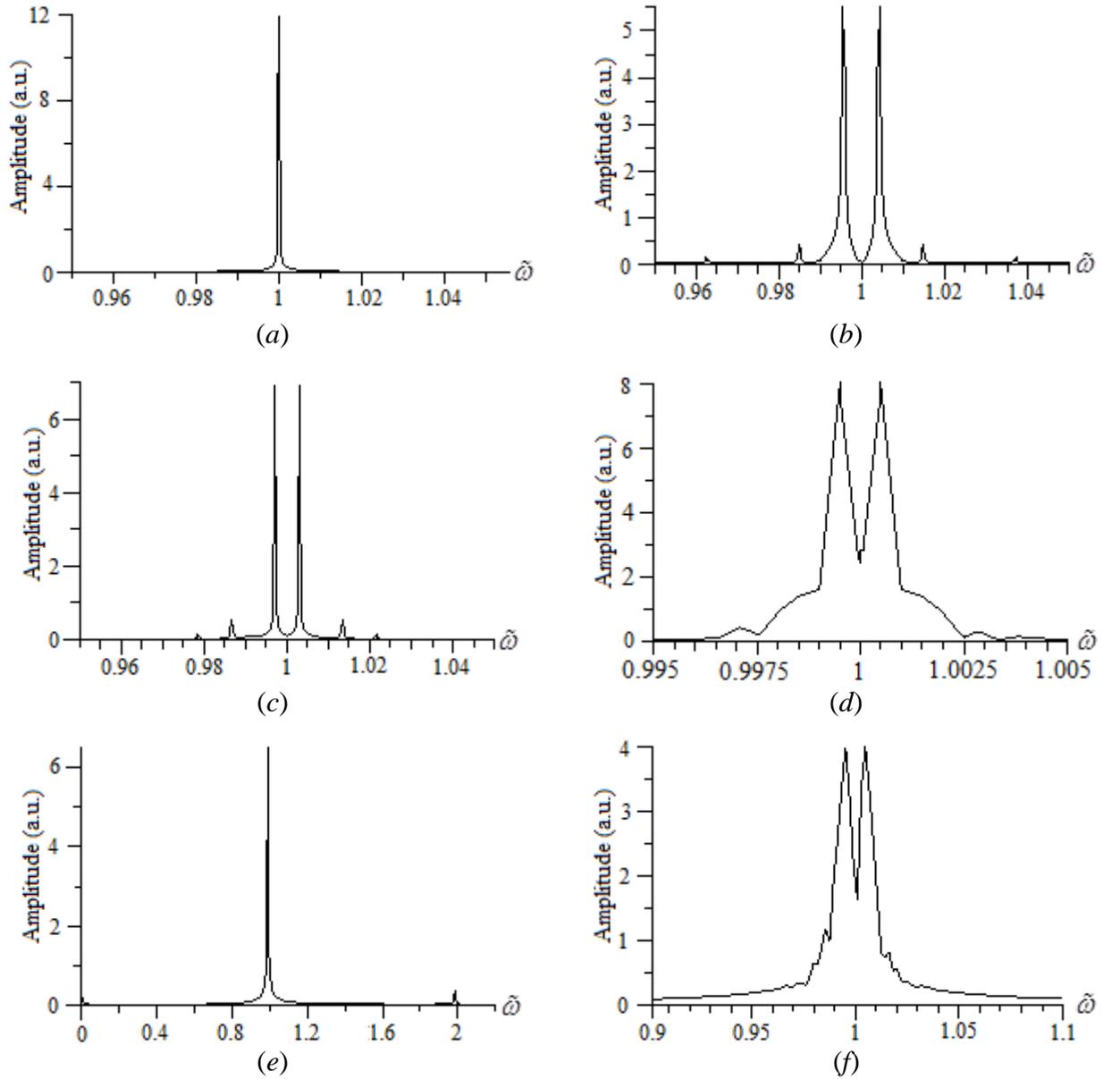

Fig.4. Frequencies in the system of two and 100 spins organized in a line at different angles relative to the field $H_0$ (amplitudes are in arbitrary units). The spectrum is centered at the Larmor frequency $\omega_0$, which in dimensionless units $\tilde{\omega} = \omega/\omega_0$ is $\tilde{\omega}_0 = 1$.

(a) line of two spins at the angle $\vartheta_{magic}$ with the $Oz$ axis (in this case secular interactions are zero), $p_d = 0.01$;

(b) line of two spins along $Oz$ axis (in this case non-secular interactions are zero), $p_d = 0.01$;

(c) line of two spins along $Ox$ axis, $p_d = 0.01$. The splitting is two times less than in (b);

(d) line of two spins along $Oz$ axis, $p_d = 0.001$ is 10 times smaller than in panel (b), correspondingly the frequency splitting is 10 times less, than in panel (b);

(e) line of 100 spins at the angle $\vartheta_{magic}$ with $Oz$ axis, $p_d = 0.05$; non-secular interactions results in weak peaks at $\omega = 0$ and $\omega = 2$;

(f) line of 100 spins along $Oz$ axis, $p_d = 0.01$.

Finally, we perform simple analytical calculations for the Pake doublet for two quantum spins of ½ to compare with the results of the classical approach above. (Similar comparisons can be provided for three or more spins.)

Evaluation of the direct products of the spin matrices gives the following expression for the Hamiltonian (in units of $\hbar\omega_0$; the azimuthal angle $\varphi=0$):

$$\hat{\mathcal{H}} = \hat{\mathcal{H}}_Z + \hat{\mathcal{H}}_{dd} = -\begin{pmatrix} 1 & 0 & 0 & 0 \\ 0 & 0 & 0 & 0 \\ 0 & 0 & 0 & 0 \\ 0 & 0 & 0 & -1 \end{pmatrix} + \frac{1}{2}\frac{\omega_d}{\omega_0}\left\{\begin{pmatrix} 1 & 0 & 0 & 0 \\ 0 & -1 & -1 & 0 \\ 0 & -1 & -1 & 0 \\ 0 & 0 & 0 & 1 \end{pmatrix}(1-3\cos^2\vartheta) - \right.$$

$$\left. -3\begin{pmatrix} 0 & 1 & 1 & 0 \\ 1 & 0 & 0 & -1 \\ 1 & 0 & 0 & -1 \\ 0 & -1 & -1 & 0 \end{pmatrix}\sin\vartheta\cos\vartheta - 3\begin{pmatrix} 0 & 0 & 0 & 1 \\ 0 & 0 & 0 & 0 \\ 0 & 0 & 0 & 0 \\ 1 & 0 & 0 & 0 \end{pmatrix}\sin^2\vartheta\right\}$$

(22)

Consider different angles between the axis connecting spins and the external field.

For $\vartheta=0$ (spins on the z-axis)

$$\hat{\mathcal{H}}_{\vartheta=0} = -\omega_d\begin{pmatrix} -1/\omega_d+1 & 0 & 0 & 0 \\ 0 & -1 & -1 & 0 \\ 0 & -1 & -1 & 0 \\ 0 & 0 & 0 & 1/\omega_d+1 \end{pmatrix}$$

(23)

and the eigenvalues of this Hamiltonian (here and below $\omega_0=1$, $\hbar=1$) are:

$$E_{1,2,3,4} = \{0, +(1-\omega_d), 2\omega_d, -(1+\omega_d)\}.$$

(24)

For $\vartheta=\pi/2$ (spins on the x-axis)

$$\hat{\mathcal{H}}_{\vartheta=\pi/2} = \frac{1}{2}\omega_d\begin{pmatrix} 1+2/\omega_d & 0 & 0 & -3 \\ 0 & -1 & -1 & 0 \\ 0 & -1 & -1 & 0 \\ -3 & 0 & 0 & 1-2/\omega_d \end{pmatrix}$$

(25)

with the eigenvalues

$$E_{1,2,3,4} = \left\{0, -\omega_d, \frac{\omega_d - \sqrt{9\omega_d^2+4}}{2}, \frac{\omega_d + \sqrt{9\omega_d^2+4}}{2}\right\}.$$

(26)

For $\omega_d \ll 1$ it gives

$$E_{1,2,3,4} = \left\{0, -\omega_d, \frac{\omega_d}{2}-1, \frac{\omega_d}{2}+1\right\}.$$

(27)

For the angle $\vartheta = \vartheta_{magic} = \cos^{-1}(1/\sqrt{3})$ the eigenvalues are $E_1 = -E_2 = -1$, $E_3 = E_4 = 0$ with the accuracy of about $\omega_d^2$ – it agrees with the absence of Pake's splitting in the classical approach above (compare with Fig.4a,e). When $\vartheta \neq \vartheta_{magic}$, the Larmor frequency $\omega_0 = 1$ splits and from formulas (24) and (27) we can conclude that the dipole interactions lead to energy separation (frequency splitting), which in case $\vartheta = \pi/2$ is half of that when $\vartheta = 0$ – same result as in the classical calculations in Fig.4b,c. Clearly, for two spins only periodic transitions between the states with different magnetic quantum numbers occur. The two spin quantum solution corresponds to the classical one both in terms of net magnetizations as well as the time dependence of the magnetizations of individual spins.

## Conclusion

The correspondence principle for dipole-interacting magnetic moments is formulated in an explicit form. Its existence in quantum mechanics is interesting by itself. It also gives the base for commonly used modeling of spin systems dynamics with operators substituted by their expectation values, bringing a complexity growth of many-body problems from exponential to polynomial. Although in this approach not all the information about the quantum systems can be studied (e.g. the transitions between quantum states) it allows an investigation of the dynamics of many significant macroscopic quantities. When there is large number of spins contributing to collective phenomena, many observables can be computed with classical equations [8-13]. The correspondence principle formulated in this paper gives a support for this approach. We also shed the light on the question how many magnetic moments are needed for reliable simulations. In the examples provided in this paper we demonstrate that the classical equations can be used for simulation of some actual spin physics problems.

## Acknowledgements

We are thankful to Drs. E.Fel'dman, G.Furman, H.Desvaux, I.Kuprov, and I.Lunegov for helpful discussions. Financial support was provided by the Russian Foundation for Basic Research (RFBR) project 13-02-96018 and Perm Ministry of Education, grant C-26/628.

## References

[1] A. Abragam, *Principles of Nuclear Magnetism*. (Clarendon Press, Oxford, 1961).
[2] S.J. Knak Jensen and O. Platz, Phys. Rev. B**7**, 31 (1973).
[3] J.H. Van Vleck, Phys. Rev. **74**, 1168 (1948).
[4] W.-K. Rhim, A. Pines and J.S. Waugh, Phys. Rev. Lett. **25**, 218 (1970).
[5] I. Kuprov, N. Wagner-Rundell, P.J. Hore, J. Magnetic Resonance, **189**, 241 (2007).
[6] I. Kuprov, private communications.
[7] G.E. Pake, J. Chem. Phys. **16**, 327 (1948).
[8] C.L. Davis, I.V. Kaganov and V.K. Henner, Phys. Rev. B**62**, 12328 (2000).
[9] C.L. Davis, V.K. Henner, A.V. Tchernatinsky and I.V. Kaganov, Phys. Rev. B**72**, 054406 (2005).
[10] V.I.Yukalov, V.K.Henner, P.V. Kharebov and E.P.Yukalova, Laser Phys. Lett. **5**, 887 (2008).
[11] V. Henner, Y. Raikher and P. Kharebov, Phys. Rev. B**84**, 144412 (2011).
[12] P.V. Kharebov, V.K. Henner and V.I. Yukalov, J. Appl. Phys. **113**, 043902 (2013).
[13] V. Henner, H. Desvaux, T. Belozerova, D. Marion, P. Kharebov, A. Klots, J. Chem. Phys. **139**, 144111 (2013).